\documentclass[aps,prl,twocolumn,showpacs,superscriptaddress,groupedaddress]{revtex4}
\usepackage{amsmath}
\usepackage{graphicx}
\usepackage{dcolumn}
\usepackage{bm}

\usepackage[utf8]{inputenc}
\usepackage[T1]{fontenc}
\usepackage{mathptmx}
\usepackage{siunitx}

\begin{document}
	
\title[Tuning superparamagnetism in CoFeB/MgO/CoFeB pMTJ]{Tuning superparamagnetism in perpendicular magnetic tunnel junctions}

%
\affiliation{Center for Spinelectronic Materials and Devices, Department of Physics, Bielefeld University, Universitaetsstrasse 25, 33615 Bielefeld, Germany}
\author{G.~Reiss} \affiliation{Center for Spinelectronic Materials and Devices, Department of Physics, Bielefeld University, Universitaetsstrasse 25, 33615 Bielefeld, Germany}
\author{J.~Ludwig} \affiliation{Center for Spinelectronic Materials and Devices, Department of Physics, Bielefeld University, Universitaetsstrasse 25, 33615 Bielefeld, Germany}
\author{K.~Rott} \affiliation{Center for Spinelectronic Materials and Devices, Department of Physics, Bielefeld University, Universitaetsstrasse 25, 33615 Bielefeld, Germany}
\email{reiss@physik.uni-bielefeld.de}
\vskip 0.25cm

\email{reiss@physik.uni-bielefeld.de}

\date{\today}

\begin{abstract}
	Thin electrodes of magnetic tunnel junctions can show superparamagnetism at surprisingly low temperature. We analysed their thermally induced switching for varying temperature, magnetic and electric field. Although the dwell times follow an Arrhenius law, they are orders of magnitude too small compared to a model of single domain activation. Including entropic effects removes this inconsistency and leads to a magnetic activation volume much smaller than that of the electrode. Comparing data for varying barrier thickness then allows to separate the impact of Zeman energy, spin-transfer-torque and voltage induced anisotropy change on the dwell times. Based on these results, we demonstrate a tuning of the switching rates by combining magnetic and electric fields, which opens a path for their application in noisy neural networks. 
	
\end{abstract}
\maketitle

Magnetic tunnel junctions (MTJs) with magnetically perpendicular CoFeB electrodes \cite{Ikeda.2010} are key components for hard disk read heads \cite{Maat.2016} and low power nonvolatile memories \cite{Wang.2013}. For such MTJs with CoFeB electrodes thinner than about $ \leq 1.5nm$, an unexpectedly low critical current density for spin-transfer-torque switching in the range of $10^4 A/cm^{2}$ has been found \cite{Leutenantsmeyer.2015}, and recent reports demonstrated a superparamagnetic behavior \cite{Parks.2018,Carboni.2019}, i.e. a thermally activated switching of the magnetic electrode that depends on the size, the temperature and a variety of external parameters. The observed switching rates, however, are not compatible with a magnetic single-domain behavior of the electrode, because the energy barrier for magnetization reversal would be by orders of magnitude too large. A granular structure of the CoFeB has been discussed \cite{Tsai.2014}, but there is no unambiguous explanation up to now.

Such superparamagnetic MTJs (sp-MTJs) can serve to study superparamagnetism \textit{life} and be useful for applications. Recently, we proposed a true random number generator based on sp-MTJs \cite{Reiss.2018}. Moreover, they can serve in noisy neural-like computing. One precondition is a pronounced maximum of the sp-MTJ’s thermal switching rate in dependence of an external input and a shift of these tuning curves by another external parameter. Mizrahi et al. \cite{Mizrahi.2018, Mizrahi.2018b} demonstrated this by varying the current through sp-MTJs.

We investigated magnetically perpendicular sp-MTJs with an exchange biased reference CoFeB electrode \cite{Manos.2019}, an MgO barrier and a \SI{1.1}{\nano \meter} thick free CoFeB electrode. The film stacks were deposited by dc and rf sputtering on a Si/SiO$_2$(50) substrate. The layer sequence was Ta(5)/ Ru(30)/ Ta(10)/ Pd(2)/ MnIr(8)/ CoFe(1)/ Ta(0.4)/ Co$_{4}$Fe$_{4}$B$_{2}$(0.8)/ MgO(X)/ Co$_{4}$Fe$_{4}$B$_{2}$(1.1)/ Ta(3)/ Ru(3) (units in \si{\nano \meter}), and the thickness X of the MgO was \SI{1.2}{\nano \meter}, \SI{1.4}{\nano \meter} or \SI{1,6}{\nano \meter}. To set the exchange bias and to crystallize the CoFeB/MgO/CoFeB, the samples were annealed at \SI{300}{\degreeCelsius} for \SI{30}{\minute} in a perpendicular magnetic field of \SI{0.7}{\tesla}. The films were patterned into circular pillars with a nominal diameter of \SI{140}{\nano \meter} and contacted via Au contacts by electron beam lithography. The overlap between the upper and lower contact was kept as small as possible to reduce the capacitive coupling and to enable a detection of the current through the MTJ at high frequency.

First, we analyzed the plane films' quasistatic magnetic properties (see supplement appendix A), which means that the typical time needed for one data point is \SIrange{0.1}{1}{\second}. When an external field is applied in-plane at room temperature, the free CoFeB layer shows an anisotropy field $H_K \approx$ \SI{330}{\kilo \ampere \per \meter}. The coercive field $H_C$ measured with an out-of-plane external field, however, is much smaller than $H_K$ and decreases strongly with temperature. Similar properties are found for the quasistatic tunneling magnetoresistance minor loops of the patterned MTJs. For the example discussed in the supplement, $H_C$ is around \SI{0.8}{\kilo \ampere \per \meter} at \SI{50}{\degreeCelsius} and reaches zero at $\approx$ \SI{85}{\degreeCelsius}. The magnetic properties of our film systems and MTJs thus are similar as reported by, e.g., Zhu et al. \cite{Zhu.2012}. 
To elucidate this puzzling switching behavior, we analyzed the time dependence of the current through MTJs with \SI{140}{\nano \meter} diameter under varying temperature, magnetic and electric field.

\begin{figure*}[t]
	\centering
	\includegraphics[width=0.8\textwidth]{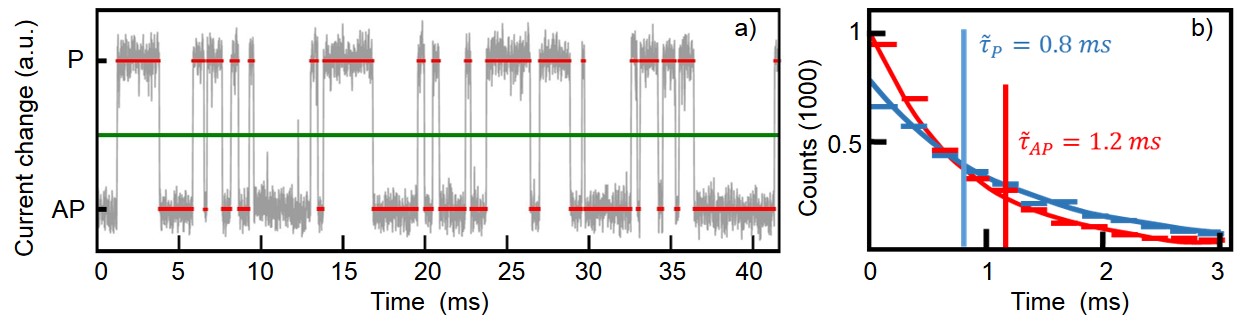}
	\caption{a) Time resolved current through an MTJ with a barrier thickness of \SI{1.4}{\nano \meter} at \SI{48}{\degreeCelsius}, \SI{-190}{Oe} magnetic bias field and \SI{100}{\milli \volt} bias voltage. The red lines indicate individual dwell times in the parallel (P) and the antiparallel (AP) state. The green line separates the two states. b) Histogram of the dwell times (P: blue, AP: red) of the complete data set with an exponential fit and the mean well times $\bar{\tau}_{P/AP}$.}
	\label{fig:Ivst}
\end{figure*}

 From the typical example shown in figure \ref{fig:Ivst} a) it becomes clear, that the MTJ is in a superparamagnetic state with a switching time in the ms-range at \SI{48}{\degreeCelsius}. The distributions of dwell times $\tau_{P/AP}$ taken from data as that of figure \ref{fig:Ivst} a) follow an exponential law as shown in figure \ref{fig:Ivst} b) with the with mean dwell times $\bar{\tau}_{P/AP}$. This property of the MTJs has severe consequences, because extrinsic quantities such as $H_C$ that are evaluated from quasistatic measurements will depend on the time scale, at which data are taken. If we, e.g., take a magnetic hysteresis loop with an out-of-plane external magnetic field at a typical measurement time much larger than $\bar{\tau}_{P/AP}$, the apparent $H_C$ will be zero, whereas non-zero values will be found otherwise. Nevertheless, the system maintains its pronounced out-of-plane anisotropy and switches only between the parallel and the antiparallel state with mean dwell times $\bar{\tau}_{P/AP}$. Thus the (intrinsic) anisotropy field $H_K$ evaluated from quasistatic characterization is still valid.

To obtain quantitative data for the superparamagnetic state of our MTJs, we have to discuss briefly the statistics of the dwell times. With perpendicular magnetic anisotropy and in a single domain approach, an Arrhenius law describes the mean dwell time $\bar{\tau}_{P/AP}$ in the Neél-Brown model \cite{Petracic.2010,Brown.1963}: $\bar{\tau}_{P/AP} = \tau_0 \cdot \exp (\Delta E_{P/AP}(\vec{H}, \vec{E}) / k_B T)$. $\Delta E_{P/AP}(\vec{H}, \vec{E})$ is the energy barrier depending on the magnetic and electric field $\vec{H}$ and $\vec{E}$, $k_B$ the Boltzmann constant. $\tau_{0}$ is the attempt time, which is the inverse of the ferromagnetic resonance frequency \cite{Wild.2017}. For ferromagnetic out-of-plane systems the attempt time at zero external field is of the order of $10^{-11} s$ \cite{Beaujour.2009}. In real superparamagnetic systems with possibly granular substructure \cite{Tsai.2014}, however, the entropy $S = k_B \cdot Ln(w)$ will play a significant role, because the system has plenty of possible pathways $w$ for magnetization switching. Using the free energy $F = E-TS = E-Tk_BLn(w)$ results in \cite{Wild.2017}:

\begin{equation}
\bar{\tau}_{P/AP} = \frac{\tau_0}{w} \exp\left(\frac{\Delta E_{P/AP}(\vec{H}, \vec{E}) }{k_B T}\right). 
 \label{eq:TaumitS}
\end{equation}

To correspondingly analyze the behavior in more detail, we thus have three handles: the temperature T, the magnetic and the electric field. 

\begin{figure}[h]
	\includegraphics[width=0.7\columnwidth]{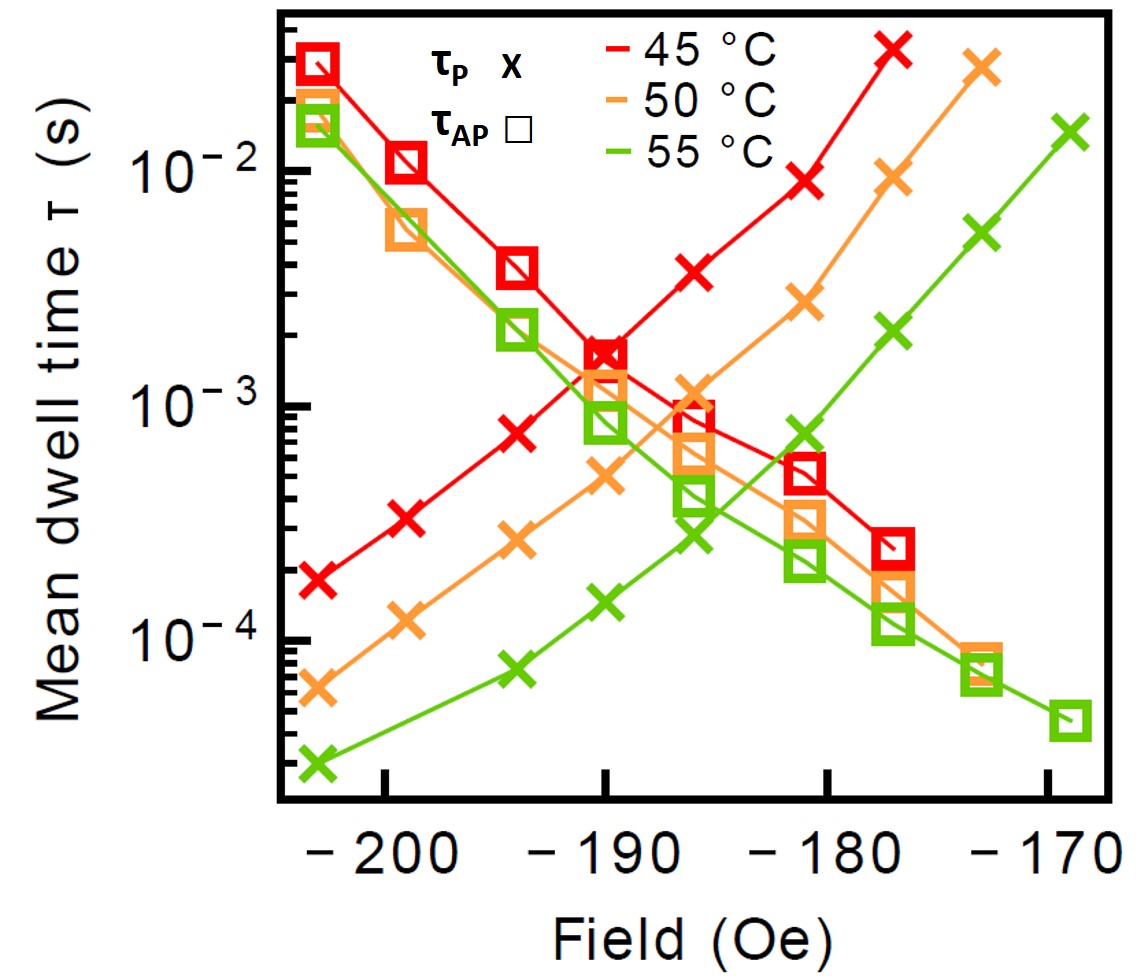}
	\caption{Mean dwell times $\bar{\tau}_{P/AP}$ as a function of the external magnetic field for varying temperature for an MTJ with $t_{MgO}=$\SI{1.4}{\nano \meter}.}
	\label{fig:Dwell14}
\end{figure}

If we apply a magnetic field $H$ perpendicular to the film plane, the Zeman energy $E_Z=\mu_0 V_{E/A} \vec{M} \cdot \vec{H}$ ($\mu_0$: magnetic vacuum permeability) will prefer an alignement of the magnetization and $H$ and thus change the mean dwell times correspondingly.  $V_{E/A}$ is either the electrode's volume $V_E$ \cite{Fruchart.1999,Kirilyuk.1997} or in case of granularity the magnetic activation volume $V_A$ \cite{Kirby.2000}. The dependence of the energy barrier in equation \ref{eq:TaumitS} on $\vec{H}$ in absence of spin torque is then given by $\Delta E_{P/AP}(\vec{H}) = \Delta E + E_Z$. In figure \ref{fig:Dwell14}, we show exemplarily the dwell times as a function of the perpendicular magnetic field for an MTJ with \SI{1.4}{\nano \meter} MgO thickness at different temperatures.

Since the pinned and the free magnetic electrode are ferromatically coupled due to their magnetic stray field, $\bar{\tau}_{P}$ and $\bar{\tau}_{AP}$ are equal at an external field $H_{comp}$ that compensates the coupling and adds an energy of $\mu_0 V_{E/A} H_{comp}M$ to the basic energy barrier $\Delta E=K \cdot V_{E/A}$ (K: effective anisotropy). We thus evaluated the data by shifting the field axis to $H' = H - H_{comp}$. With this and the collinearity of $\vec{M}$ and $\vec{H}$,  $\bar{\tau}_{P/AP}$ are given by:

\begin{equation}
\bar{\tau} (H,T)  = \frac{\tau_0}{w} \exp \left(\frac{\Delta E + \mu_0 V_{E/A} M_S H'}{k_B T}\right).
\label{eq:tau(H,T)}
\end{equation}

The energy barrier $\Delta E$ and the product $V_{E/A} M_S$ can be determined from taking the derivative of $Ln(\bar{\tau} / \tau_0) $ with respect to $1/k_B T$ or $\mu_0 H$, respectively. The results of this evaluation are summarized in table \ref{Table1}.

\begin{center}
	\begin{table}[h]
	\begin{tabular}{|c c c c c c|} 
		\hline
		$t_{MgO}$ & $V_{E/A} M_S$ & $\Delta E$ & $K^*$ & $K / K^*$ & $M_S $\\ [0.5ex] 
		$nm$ & $A nm^2$ & $eV$ & $kJ/m^3$ &   & $kA/m$\\
		\hline\hline
		1.2 & 9,28 & 2.5 & 26 & 12.7 & 1022\\ 
		\hline
		1.4 & 8,33 & 1.3 & 13,5 & 24,3 & 918\\
		\hline
		1.6 & 6,29 & 2.4 & 24,9 & 13,2 & 693\\
		\hline
	\end{tabular}
	\caption{The values for the saturation magnetic moment $V_{E/A} M_S$, the activation energy $\Delta E$, the apparent effective anisotropy $K^*=\Delta E / V_E$, the ratio $K/K^*$ for three MgO thicknesses with the MTJ's electrode volume $V_E=1.54 \cdot 10^{-23}m^3$, and the saturation magnetization $M_S$ evaluated with the magnetic activation volume $V_A$.}
	\label{Table1}
\end{table}
\end{center}

If we compare the measured effective anisotropy ($K \approx 330 kJ/m^3$) with the apparent value $K^*$ deduced from $K^*=\Delta E / V_E$, there is a large discrepancy ranging from $K/K^*=12.7$ to $24.3$ with a mean value of $16.7 \pm 7$. Taking, however, granularity of the electrode into account, then the magnetic activation volume $V_A = \Delta E / K$ is correspondingly smaller. With a radius of the electrode of $r_E=$\SI{70}{\nano \meter}, the radius of the activation volume is $r_A=r_E / \sqrt{K / K^*} = (17 \pm 4)$nm . One test of this model is the resulting saturation magnetization $M_S$, which was reported to be between between \SI{500}{\kilo \ampere / \meter} and \SI{1}{\kilo \ampere / \meter} \cite{Tsai.2014,Manos.2019} depending on the preparation conditions and if dead magnetic layers at the interfaces are considered or not. Using the full electrode's volume, we obtain from $V_E M_S$ in table \ref{Table1} an $M_S^*$ smaller than \SI{250}{\kilo \ampere \per \meter}, i.e. unphysically low values. Using the determined magnetic activation volume $V_A$, the resulting values for $M_S$ (see table \ref{Table1}) are in agreement with the literature. Moreover, the lateral size of $V_A$ is very well comparable to the typical domain wall width found for similar CoFeB films with perpendicular magnetic anisotropy \cite{Yamanouchi.2011}, and about twice the value of the typical lateral grain size in our samples (see supplement appendix C). Thus modelling the properties with this activation volume leads to physical consistency.

Using these results, we can now estimate, how much entropic effects impact the dwell times. The effect of the entropy $S=k_B Ln(w)$ can be evaluated for $H'=0$ and small bias voltage from the intercept when plotting $Ln(\bar{\tau}_{P/AP} / \tau_{0})$ versus $(k_B T)^{-1}$. This gives values for $Ln(w)$ of the order of 35, which means that $w$ is of the order of $10^{15}$! This number seems to be extraordinarily large. Similar values have, however, already been found for the decay of skyrmionic magnetization patterns \cite{Wild.2017} and their magnitude was related with the vast amount of pathes $w_S$ for skyrmion decay. For our MTJs, we can again estimate the magnetic activation volume from the number w of the entropic pathways that can lead to their thermally activated magnetization switching. If we assume, that the switching can start at either of N sub-volumes of the free electrode, then we have around $w \approx N!$ possible pathes. Thus $\ln w \approx 35 = Ln(N!) = \sum_{i=1}^{N}Ln(i)$ resulting in $N\approx 17$ for our sp-MTJs with \SI{70}{\nano \meter} radius. The radius of a single activation volume is then $r_A=70nm / \sqrt{17}\approx 17nm$. It is remarkable, that the two approaches to evaluate $V_A$ lead to almost the same radii, although the underlying physics is in the first case the evaluation of the energetics of the switching process, while thermodynamic considerations are used in the second.

We now turn to the influence of the bias voltage on the switching process and the dwell time. If a bias voltage $U_B$ is applied to the MTJ, two additional effects act on the magnetization: first, the spin torque due to the spin polarization of the current $I=U_{B}/R_0 \cdot \exp (-B \cdot t_{MgO})$ with the MTJ's contact resistance $R_0$ and the inverse decay length B. Second, the electric field $E=U_{B} / t_{MgO}$ at the interfaces modifies the interfacial magnetic anisotropy energy density \cite{Weisheit.2007,Kanai.2012,Wang.2011} and leads to a linear change of the anisotropy with $U_{B}$ \cite{Endo.2010,Niranjan.2010,Shiota.2011}: $\Delta K (U_{B}) = \beta |\vec{E}| = \beta  U_{B}/ t_{MgO}$, where $\beta$ characterizes the strength of the dependence of $K$ on $E$. Spin torque and anisotropy change, however, have fundamentally different impact on $\bar{\tau}$: While spin torque will stabilize one of the (P, AP) states and destabilize the other one, the anisotropy change will either de- or increase both $\bar{\tau}_P$ and $\bar{\tau}_{AP}$, depending on the polarity of $U_B$. In addition, for fixed $U_{B}$, the influence of the spin torque on the magnetization decreases exponentially with increasing MgO barrier thickness. In contrast, the anisotropy change decreases only with $t_{MgO}^{-1}$. Thus for thin barriers, the spin torque can be expected to be dominant, while for thick barriers the change of the anisotropy can have larger impact.

At fixed $t_{MgO}$, the dependence of $\Delta E$ on the bias voltage is taken into account by multiplying equation \ref{eq:tau(H,T)} with the factor $\exp ((\beta' V \pm A)U_{B}/(k_B T))$, where $\beta'$ and $A$ describe the strength of the anisotropy change and the spin torque, respectively. By convention, we use the plus sign for the P- and the minus sign for the AP-state. Then, the influences of the anisotropy change and the spin torque can be separated by evaluating 

\begin{equation}
\Delta_{\pm} = \frac{k_B T}{2} \cdot \left( \frac{d}{dU}\left( Ln(\tau_{P}/s) \pm Ln(\tau_{AP}/s)\right) \right)
\label{Deltapm}
\end{equation}

where $\Delta_{+}$ gives the anisotropy change and $\Delta_{-}$ the spin torque influence (see section Data evaluation).

\begin{figure}
	\centering
	\includegraphics[width=\columnwidth]{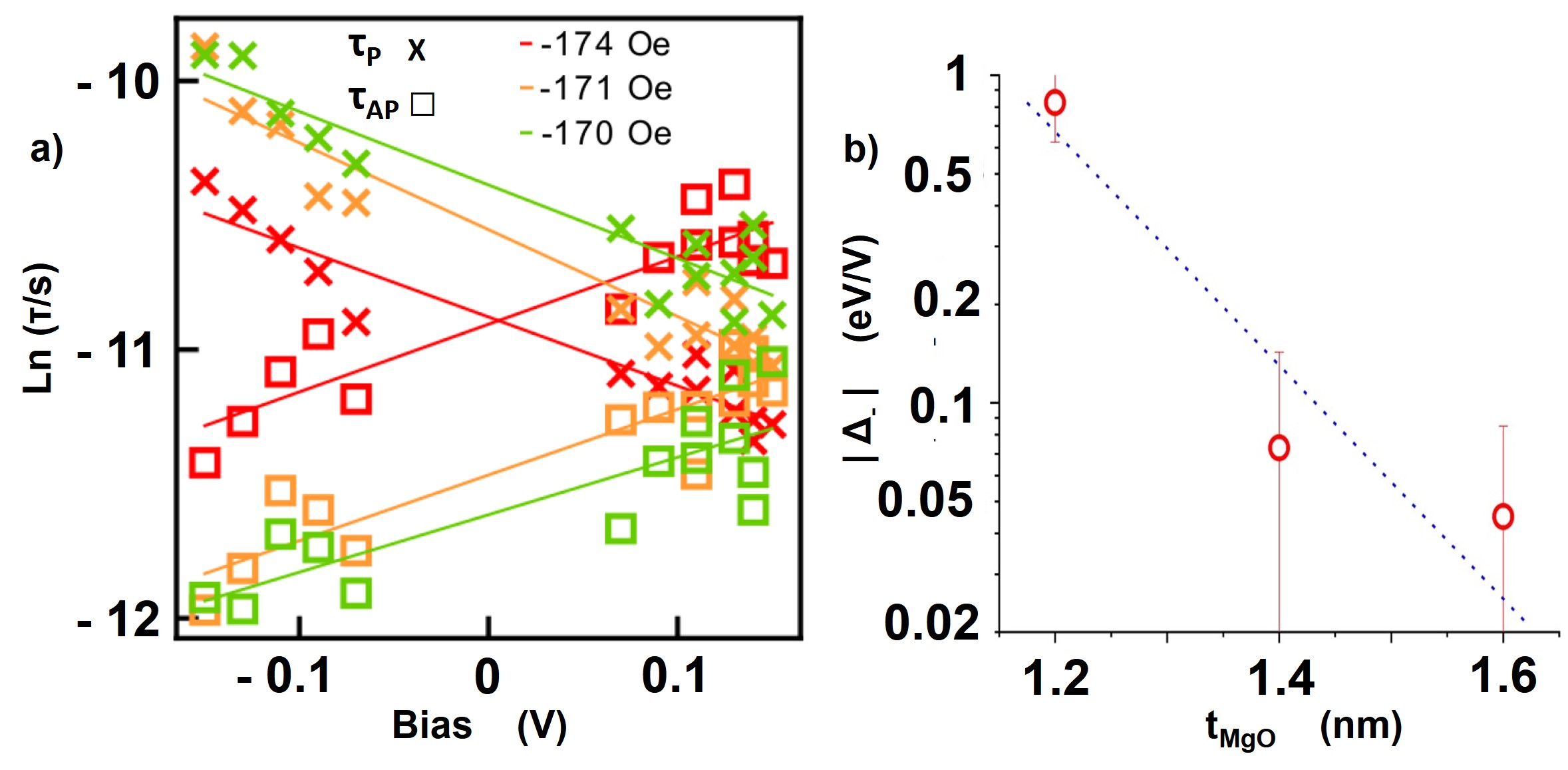}
	\caption{a) $Ln(\tau_{P/AP}/s)$ with linearized fit in dependence of $U_{B}$, each for five different magnetic fields, \SI{65}{\degreeCelsius}, on a \SI{140}{\nano \meter} structure and MgO thicknesses of \SI{1.4}{\nano \meter}. b) $|\Delta_-|$ in dependence of the MgO-thickness and numerical fit of the data with $|\Delta_-| = A \cdot \exp{ \left(-B \cdot t_{MgO} \right)}$.}
	\label{fig:feldkonstbiasfit}
\end{figure}

In figure \ref{fig:feldkonstbiasfit} a), we show the results for the dwell times as a function of $U_B$ for $t_{MgO}=1.2nm$ at \SI{65}{\degreeCelsius}, and in b) the results of fitting the spin torque part $\Delta_{-}$ as a function of $t_{MgO}$ with an exponential function. As expected, the spin torque term is for \SI{1.2}{\nano \meter} barrier thickness with $\approx$ \SI{0.83}{\electronvolt / \volt}  more than four times larger than that of the anisotropy change ($\approx$ \SI{0.17}{\electronvolt / \volt}), while for $t_{MgO}=1.6nm$ this ratio reduces to two ($\left| \Delta \right|_-\approx$ \SI{0.045}{\electronvolt / \volt} and $\left| \Delta \right|_+\approx$ \SI{0.021}{\electronvolt / \volt}). The numerical evaluation of the data for the three MgO thicknesses gives a spin torque term $\Delta_- = -(20 \pm 2) \frac{pJ}{V} \times \exp \left(-(13 \pm 6)/nm \cdot t_{MgO} \right)$, which mirrors the exponential dependence of the current density on $t_{MgO}$. The anisotropy contribution results in $\beta = \beta' \cdot t_{MgO} = (30 \pm 15) fJ/Vm$. The separation of the small anisotropy contribution from the spin torque term is, however, not very reliable. The value obtained here is, however, very similar to the result of $-33 fJ/VM$ obtained by Endo et al. for MgO/CoFeB/Ta thin films \cite{Endo.2010}, which corresponds to the free layer system used in our MTJs.

Now, we can combine the impacts of the magnetic field and the electric field to realize tuning curves. If an sp-MTJ is only either in the P or the AP state such as the 140nm diameter devices used in this work, the tuning curve is given by the switching frequency $\nu (H, U_B) = 2 / (\bar{\tau}_P(H, U_B) + \bar{\tau}_{AP}(H, U_B))$. If the barrier thickness is too large, the application of $U_B$ would change only the anisotropy. This would give a simultaneous de- or increase of both dwell times and thus tuning of the switching rates, i.e. a shift of the gaussian dependence on one parameter by the other one would not be possible. If, however, spin torque is dominating the impact of $U_B$, one of the dwell times is driven exponentially to zero and the other one to infinity by application of either $H$ (figure \ref{fig:Dwell14}), or $U_B$ (figure \ref{fig:feldkonstbiasfit}). In the corresponding range of barrier thickness, we can, therefore, use the combination of both to shift the dependence of the switching rate on one parameter by varying the other one. 

\begin{figure}[h]
	\centering
	\includegraphics[width=0.9\columnwidth] {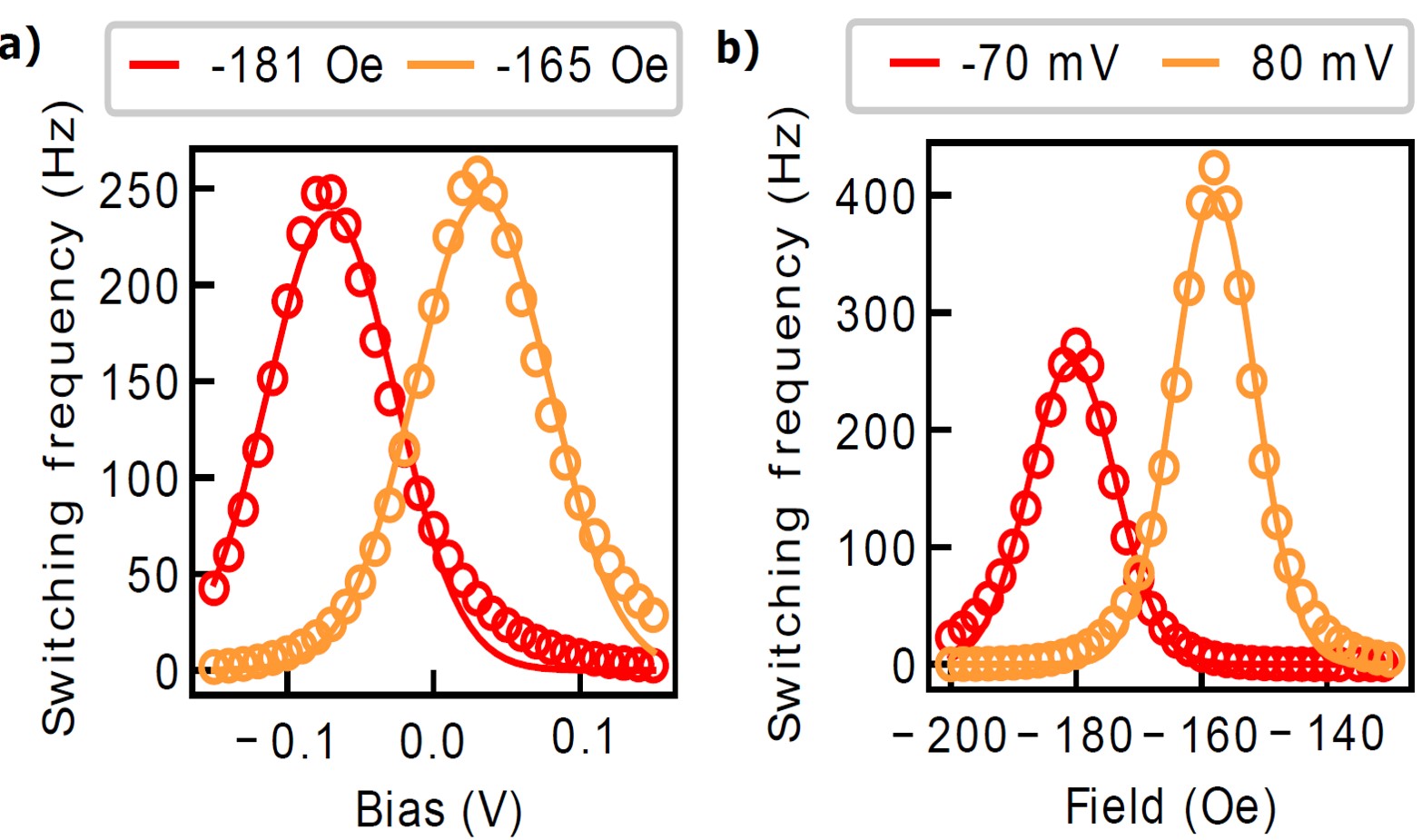}
	\caption{Tuning curves for a \SI{1.2}{\nano \meter} MgO thick barrier and \SI{140}{\nano \meter } structure at \SI{65}{\degreeCelsius} (circles) and fit with a Gaussian function (lines). a) Bias dependence with constant magnetic field. b) Magnetic field dependence with constant bias voltage. }
	\label{fig:switchingfreq}
\end{figure}

The switching rate $\nu (H, U_B)$ was evaluated exemplarily for the sample with \SI{1.2}{\nano\meter} thick MgO barrier in dependence of the two input parameters bias voltage and magnetic field. The switching frequency tuning curves as well as Gaussian fits are shown in figure \ref{fig:switchingfreq} for constant magnetic field depending on the bias voltage (figure \ref{fig:switchingfreq} \textbf{a)}) and for constant bias voltage depending on the magnetic field (figure \ref{fig:switchingfreq} \textbf{b)}).

The results in figure \ref{fig:switchingfreq} are in agreement with our finding, that spin torque dominates the impact of $U_B$ for a the \SI{1.2}{\nano \meter} thick tunneling barrier. The position of the peak for one input parameter can be easily shifted by varying the other one. The asymmetry in figure \ref{fig:switchingfreq} \textbf{b)} comes from the change of the anisotropy, which de- or increases the switching rates depending on the sign of the applied voltage $U_B$.

The tuning curves can be very well described with a Gaussian dependence on both input parameters $H$ and $U_B$:

\begin{equation}
\nu=\nu_0 \cdot \exp \left(\frac{1}{2}\left(\frac{C-C_0}{\sigma}\right)^2  \right)
\label{nuvsH}
\end{equation}

where $C$ is either the magnetic field or the bias voltage, $C_0$ is the peak position and $\sigma$ the full width at half maximum.

For, e.g., figure \ref{fig:switchingfreq} b, we find $H_0= 157.8 Oe$, $\sigma=6.29 Oe$ and $H_0= 180.6 Oe$, $\sigma=6.95 Oe$ for for $U_B = 80mV$ and for $U_B = -70mV$, respectively. This tuning of the switching rates matches exactly the requirements for the firing rates of neurons in population coding networks \cite{Salinas.1995} and thus can emulate tuning curves for noisy neural-like computing\cite{Mizrahi.2018,Pouget.2000,Mizrahi.2018b}.

\begin{acknowledgments}

The authors gratefully acknowledge the support of the work by the Deutsche Forschungsgemeinschaft under contract RE 1052/22-1 and -2. We also thank Hans-Werner Schumacher (PTB Braunschweig) for Kerr-microscopy.

\end{acknowledgments}

\bibliographystyle{unsrt}

\bibliography{literatur}

\end{document}